# Localized excitations and scattering of spin waves in ferromagnetic chains containing magnetic nanoclusters

## O.V. Charkina, M.M. Bogdan

*B. Verkin Institute for Low Temperature Physics and Engineering of the National Academy of Sciences of Ukraine, 47 Nauky Ave., Kharkiv, 61103, Ukraine*

E-mail*:* charkina@ilt.kharkov.ua, m_m_bogdan@ukr.net

The spectrum of localized excitations in an anisotropic one-dimensional ferromagnet containing a spin cluster of arbitrary size was found accurately within the framework of the discrete Takeno-Homma model. The stability boundaries of spin nanoclusters are determined depending on their size and the ferromagnet's exchange and anisotropy parameters. The problem of scattering of spin waves by nanoclusters is solved and explicit analytical expressions are obtained for their reflection and transmission coefficients. A model of a metamaterial consisting of weakly interacting magnetic molecular nanoclusters with the discovered dynamic properties is proposed.

## 1. Introduction

Nanocluster magnetic structures in quasi-one-dimensional Ising ferromagnets were discovered more than half a century ago in experiments on the resonant absorption of a high-frequency electromagnetic field. This phenomenon was called spin-cluster resonance [1]. The spin cluster is a nanodomain consisting of several spins oriented opposite to the matrix spin direction. In quasi-one-dimensional ferromagnets, the spin cluster is limited by two Ising domain boundaries; therefore, the nanocluster energy does not depend on its size. The existence of such magnetic nanostructures is also possible in systems described by Heisenberg spin chains model with similar exchange interactions and easy-axis magnetic anisotropy [2,3]. Quasi-one-dimensional ferromagnets possessing such properties at low temperatures are organometallic compounds of the TMNB, TMNC, TMANC, and FeTAC type [4-6]. The existence of discrete domain boundaries of the Ising type in such crystals is evidenced by magnetic resonance experiments on the detection of internal oscillation modes, which demonstrates the presence of the local absorption peaks at frequencies below the edge of the spinwave spectrum [5].

In recent years, a whole class of new magnetic nanoobjects has been synthesized [7], the so-called "magnetic molecules," which have total spin of the order of a dozen Bohr magnetons.

Such molecular nanoclusters consist of alternating magnetic ions and radicals whose spins are not the same; therefore, in the fundamental state, they are ferrimagnets [8]. At low temperatures, these nanoclusters form a molecular crystal in which the "molecule" moments interact weakly with each other. Arranging such "magnetic molecules" in the crystal in a regular order in the form of chains separated by nonmagnetic organic structures, one may obtain quasi-one-dimensional metamaterial with given structure period and, accordingly, with given interaction parameter between classical spins, which can be either of the same order or much less than the single-ion anisotropy constant. In such a metamaterial, spin clusters formed by "magnetic molecules" are actually mesoscopic objects. It is obvious that much simpler control of their magnetic orientation by the local fields compared to the orientation of individual spins makes these structures very attractive from the point of view of their use as basic memory elements in new computer technologies.

In this context, an important circumstance is that the stability of the cluster essentially depends on the number of spins in it and the overall magnetic resonance properties of the quasi-one-dimensional ferromagnet containing it. Therefore, theoretical analysis of the spectrum of weak excitations of anisotropic Heisenberg chains containing spin nanoclusters is of great interest. The main purpose of this paper is to investigate discreteness effects due to complete consideration of exchange interaction in the motion equations for a strongly anisotropic ferromagnet. The description of the dynamics is carried out within the framework of the Takeno-Homma equation [9] for the azimuthal angles of the node spins, to which the discrete Landau-Lifshitz equations are resolved for a ferromagnet with anisotropic light plane and exchange close to the anisotropy constant of the easiest magnetization axis [10]. For such a system, exact solutions describing nanoclusters are considered in this paper, and the structure of the entire spectrum of their small excitations is analyzed. The spectrum of localized excitations of a ferromagnet containing nanoclusters is precisely found, and their stability limits are determined. It is shown that instability leads either to the destruction of nanoclusters, or to their transformation into magnetic domains bounded by noncollinear domain walls. To analyze singularities of the continuous-spectrum waves dynamics, the problem of spin waves scattering on arbitrary size clusters is exactly solved, and explicit expressions for their reflection and transmission coefficients are obtained.

To analyze singularities of the continuous-spectrum waves dynamics, the problem of spin waves scattering on arbitrary size clusters is exactly solved, and explicit expressions for their reflection and transmission coefficients are obtained. of self-localization of nonlinear oscillations [13], which, in a highly discrete case, represent inhomogeneities similar to local distortions in a crystal. Therefore, it is possible to use methods based on the theory of defects when studying the influence of discrete solitons on crystal dynamics. Using approaches developed by Kosevich, we remember with deep gratitude our teacher, and dedicate this paper to his memory on the 90th anniversary of his birth.

## 2. The Takeno-Homma model for an anisotropic ferromagnet

The classical Heisenberg model of a ferromagnetic chain with biaxial single-ion anisotropy has been used in numerous studies to interpret experimental results [4-6] and to describe nonlinear effects in the theory of magnetism [14]. The Hamiltonian of this model has the form

$$\mathsf{H} = -J\sum_n \mathbf{S}_n \mathbf{S}_{n+1} + \frac{1}{2}\sum_n \left( D(S_n^z)^2 - A(S_n^x)^2 \right) \qquad (1)$$

where $\mathbf{S}_n$ is the classical spin on the node with number $n$, $J$ is the exchange interaction constant, $A$ and $D$ are constants of easy-axis and easy-plane anisotropy, respectively. In an easy-axis ferromagnet, i.e., at $D=0$, discrete collinear $180^0$ domain walls and the spectrum of their localized oscillations was studied in Ref. 3. It was found the critical exchange value $J_0 = 0.75A$, at which a loss of stability of the Ising type discrete boundary occurs. Dependencies of the frequencies of the internal modes of the oscillations of the domain walls on the magnitude of the exchange interaction were also calculated. In the case of strong easy-plane anisotropy $D \gg A$, when it is assumed that only small deviation of the $\mathbf{S}_n$ vector from the easy plane is possible, the Hamiltonian (1) can be approximately reduced (see, for example, Ref. 10) to the Hamiltonian of the Takeno-Homma model [9]

$$\mathrm{H} = \frac{\hbar^2}{2DS_0^2}\sum_{n=1}^{N}\dot{\varphi}_n^2 - J\sum_{n=1}^{N}\cos(\varphi_n - \varphi_{n-1}) - \frac{1}{2}A\sum_{n=1}^{N}\cos^2(\varphi_n) \qquad (2)$$

Such a model is described by one independent scalar variable—the azimuthal angle $\varphi_n$ of the node spin $\mathbf{S}_n$. In the representation of the Hamiltonian, the dot denotes time differentiation, and S0 is the spin value. In the general case, the Takeno-Homma model takes into account the Zeeman contribution of the external magnetic field. However, in this paper, we confine ourselves to the case of the absence of a field. Then, the Lagrange motion equation corresponding to the model (2) may be represented in a dimensionless form, as follows:

$$\frac{d^2\varphi_n}{dt^2} + \lambda\big(\sin(\varphi_n - \varphi_{n-1}) - \sin(\varphi_{n+1} - \varphi_n)\big) + \cos(\varphi_n)\sin(\varphi_n) = 0. \qquad (3)$$

It is written by introducing a dimensionless exchange parameter $\lambda = J/A$ and time unit $t_0 = \hbar/(S_0\sqrt{DA})$. This Takeno-Homma equation (THE), according to the terminology used by the authors, is the $\pi$-sine-lattice equation [9], emphasizing the existence of $\pi$-solitons in its $180^0$ domain walls Equation (3) was investigated by its authors from the point of view of its similarity to integrable equations. A study of the dynamics of its nonlinear excitations using an analytical method and numerical modeling for large values of the parameter $\lambda \geq 1$ showed that the behavior of solitons and antisolitons (domain boundaries of different signs) in collisions in THE is similar to that in exactly solvable models [9]. For large exchanges ($\lambda \gg 1$), in order to maintain the same order of magnitude of all terms in THE, the sinus difference must be replaced by the difference of their arguments, i.e., second differential. In this case, Eq. (3) becomes a discrete sine-Gordon equation (DSGE)

$$\frac{d^2\varphi_n}{dt^2} + \lambda(2\varphi_n - \varphi_{n-1} - \varphi_{n+1}) + \sin(\varphi_n)\cos(\varphi_n) = 0 \qquad (4)$$

A further increase in the parameter $\lambda$ allows us to pass to the long-wave limit and to obtain exactly integrable continuum sine-Gordon equation

$$\frac{\partial^2\varphi}{\partial t^2} - \frac{\partial^2\varphi}{\partial z^2} + \sin\varphi\cos\varphi = 0, \qquad (5)$$

in which a continuous coordinate is introduced $z = n/\sqrt{\lambda}$.

As noted by Takeno and Homma, the advantage of Eq. (3) over Eq. (4) is the complete consideration of the exchange interaction, which leads to the presence of exact static solutions of the THE. They correspond to spin clusters bounded by Ising boundaries. For a cluster in an infinite chain, enclosed between nodes $j$ and $l$, the distribution of the azimuthal angle has the form

$$\varphi_n^0 = 0 \quad n < j; \qquad \varphi_n^0 = \pi \quad j \leq n \leq l; \qquad \varphi_n^0 = 0, 2\pi \quad n > l. \qquad (6)$$

The choice of the zero value for $n \to \infty$ seems to be natural for the cluster, and choice of the azimuthal angle value equal to $2\pi$ corresponds to the solution in the form of a $360^0$ domain boundary. For magnetic applications, both distributions in (6) are completely equivalent as all spins outside the cluster are in the same ground state. However, for other physical systems described by THE, it may imply particles being in different vacuum states. Then, the first configuration corresponds to a soliton-anti-soliton pair, and the second is a sequence of two identical discrete solitons. As shown below, in the case of loss of stability by the structures (6), degeneracy between them is also removed in a ferromagnet (2) as the noncollinear distributions of spins that differ from each other and have different dynamic properties arise in a bifurcational manner.

It is obvious that the energy of the ferromagnet (1) with cluster (6) based on the ground state is equal to the sum of the energies of two Ising boundaries and does not depend on the cluster size $E = 4JS_0^2$. Note that there are no exact collinear solutions in DSGE (4) and the sine-Gordon equation (5).

In Ref. 10, the stability and a spectrum of small oscillations for $180^0$ collinear and noncollinear domain walls and cluster consisting of one spin were studied in the framework of the Takeno-Homma equation (3). In this paper, we study the stability and excitation spectrum of arbitrary-size clusters (6). The results of Ref. 10 correspond to two extreme cases of solving this spectral problem for an infinite-sized cluster with virtually remote domain boundaries, and for a cluster of minimum size. In the following section, the problem of localized vibrations of a spin cluster of a THE of arbitrary size is completely solved analytically.

### 3. Spectral problem for the vibrational modes of spin clusters

Small oscillations of nanoclusters are described by Eq. (3), linearized near solution (6). Assuming additions to the cluster solution $\Delta\varphi_n(t) = \varphi_n(t) - \varphi_n^0 \ll 1$ and separating apparent time dependence $\Delta\varphi_n(t) = \psi_n \exp(i\omega t)$, we obtain the following system of algebraic equations for amplitudes $\psi_n$:

$$\lambda[(\psi_n - \psi_{n-1})\cos(\varphi_n^0 - \varphi_{n-1}^0) - (\psi_{n+1} - \psi_n)\cos(\varphi_{n+1}^0 - \varphi_n^0)] + \cos(2\varphi_n^0)\psi_n = \omega^2 \psi_n. \quad (7)$$

This system of equations is an eigenvalue problem for the squared frequency parameter $\varepsilon = \omega^2$. Positive values of e correspond to vibrational modes, and negative values of $\varepsilon = -v^2$ correspond to the modes of instability leading to exponential growth of the additives $\Delta\varphi_n(t) = \psi_n \exp(vt)$.

Let us substitute the distribution of the angles (6) into Eq. (7). Dividing it by $\lambda$ and introducing the notation

$$\beta = \frac{1-\omega^2}{\lambda}, \quad (8)$$

we obtain the spectral problem for eigenvalues of the parameter $\beta$. The corresponding system of algebraic equations can be rewritten in the following form:

$$2\psi_n - \psi_{n-1} - \psi_{n+1} + \beta\psi_n = 2\big((\delta_{n,j-1} + \delta_{n,j} + \delta_{n,l} + \delta_{n,l+1})\psi_n - (\delta_{n,j} + \delta_{n,l+1})\psi_{n-1} - (\delta_{n,j-1} + \delta_{n,l})\psi_{n+1}\big), \quad (9)$$

where indices $j$ and $l$, as follows from the solution (6), correspond to the boundary numbers of spins in the cluster. It follows from (8) and (9) that squared frequencies of all modes depend linearly on the parameter $\lambda$

$$\omega^2 = 1 - \beta\lambda, \quad (10)$$

where factors $\beta$, as eigenvalues for the spectral problem (9), depend only on the total number of spins in the system $N$ and number of spins in the cluster $m = l - j + 1$. It is obvious that localized oscillations correspond to positive $\beta$ and, respectively, $\omega^2 < 1$, and delocalized oscillations to negative values of $\beta \equiv -\mu < 0$ and $\omega^2 \geq 1$. The waves of the continuous spectrum in the limit, when the number of spins $N$ tends to infinity, with $n \to \infty$ are described by the asymptotics $\psi_n \propto \exp(ikn)$. Then, Eq. (9), taken in the region far from the cluster, implies relation between parameter $\mu$ and quasi-wave vector $k$, and, accordingly, the standard expression for the spin-waves dispersion law

$$\mu(k) = 4\sin^2\frac{k}{2}, \qquad \omega^2(k) = 1 + 4\lambda\sin^2\frac{k}{2}. \qquad (11)$$

The form of the spectral problem equation given in (9) is convenient for applying the method of local perturbations of Lifshitz [11]. Let us use it to find frequencies corresponding to internal vibrational modes of the spin cluster. First, let us find the relation between parameter $\beta$ and the decay decrement of the amplitudes of the internal modes $\kappa$, which decrease as $\psi_n \propto \exp(-\kappa n)$ when $n \to \infty$. Let us substitute this asymptotics for $\psi_n$ into Eq. (9) in the range of values $n \gg l, j$, when its right part becomes zero, and we obtain

$$\beta = 4\,\text{sh}^2\frac{\kappa}{2} \qquad (12)$$

Let us apply the Fourier transform to Eq. (9) [11]

$$\Psi_k = \sum_n \psi_n \exp(-ikn), \qquad \psi_n = \frac{1}{2\pi}\int_{-\pi}^{\pi}\Psi_k \exp(ikn)dk. \qquad (13)$$

Then, for the variable $\Psi_k$, we obtain the expression

$$\Psi_k = \frac{2}{\beta + 2(1-\cos k)}\big((1-\exp(ik))(\psi_j - \psi_{j-1})\exp(-ikj) + \\ + (1-\exp(-ik))(\psi_l - \psi_{l+1})\exp(-ikl)\big). \qquad (14)$$

Applying the inverse transformation, we obtain a system of four linear algebraic Lifshitz equations for the amplitudes $\psi_{j-1}, \psi_j, \psi_l, \psi_{l+1}$

$$(1-I_0)\psi_{j-1} + I_0\psi_j - I_{m+1}\psi_l + I_{m+1}\psi_{l+1} = 0,$$

$$I_0^*\psi_{j-1} + (1-I_0^*)\psi_j - I_m\psi_l + I_m\psi_{l+1} = 0,$$

$$I_m^*\psi_{j-1} - I_m^*\psi_j + (1-I_0)\psi_l + I_0\psi_{l+1} = 0, \qquad (15)$$

$$I_{m+1}^*\psi_{j-1} - I_{m+1}^*\psi_j + I_0^*\psi_l + (1-I_0^*)\psi_{l+1} = 0.$$

Integral coefficients $I_\gamma$, where $\gamma = 0, m, m+1$, are determined by the formula

$$I_\gamma = \frac{1}{\pi}\int_{-\infty}^{\infty}\frac{1-\exp(-ik)}{\beta + 2(1-\cos k)}\exp(-ik\gamma)dk. \qquad (16)$$

Solvability condition for the system (15) is zero determinant value of the matrix composed of the coefficients at the amplitudes $\psi_{j-1}, \psi_j, \psi_l, \psi_{l+1}$

$$D(\beta) = \left(1 - I_0 - I_0^*\right)^2 - \left|I_{m+1} - I_m\right|^2 = 0. \tag{17}$$

It is easy to determine explicit expressions for the coefficient-integrals and to verify the real type of corresponding expressions

$$I_0 = I_0^* = 1 - \sqrt{\frac{\beta}{\beta+4}}, \qquad \Delta = \Delta^* = I_{m+1} - I_m = 2\sqrt{\frac{\beta}{\beta+4}}\left[\frac{1}{2}\left(\sqrt{\beta+4} - \sqrt{\beta}\right)\right]^{2m} \tag{18}$$

Using (17) and (18), to find frequencies of the internal modes, we obtain two equations

$$1 - 2I_0 = \pm \Delta. \tag{19}$$

Let us introduce the notation $x = \exp(-\kappa) < 1$; then, from (12), we obtain the following expression for the $\beta$ parameter

$$\beta = \frac{(1-x)^2}{x}. \tag{20}$$

Let us use this substitution and reduce (19) to the following two equations:

$$\frac{1}{1-x} - \frac{3}{2} = \pm x^m. \tag{21}$$

Using graphical analysis, the existence (21) of a single positive root less than 1 in each of equations can easily be verified. Therefore, in a ferromagnet with a cluster of any size, only two localized oscillations exist.

In the case $m=1$ and $m=2$ solutions of Eq. (21) may be found in explicit form. At $m=1$, the positive roots $x_\pm < 1$ of the quadratic equation (21), and, respectively, $\beta_\pm$, are as follows:

$$x_- = \frac{1}{4}\left(5 - \sqrt{17}\right), \quad \beta_- = \frac{1}{4}\left(7 + \sqrt{17}\right); \qquad x_+ = \frac{1}{2}, \quad \beta_+ = \frac{1}{2}; \tag{22}$$

and the frequencies of internal modes $\omega_\pm = \sqrt{1 - \beta_\pm \lambda}$ are zeroed with parameter $\lambda$ values

$$\lambda_- = \frac{1}{8}\left(7 - \sqrt{17}\right), \qquad \lambda_+ = 2. \tag{23}$$

These values completely coincide with the results of the previously solved problem for a nanocluster consisting of one spin [10]. Thus, the single-spin cluster loses stability when $\lambda_- \cong 0.3596$

At $m=2$, the positive roots of cubic equations less than 1 are equal to

$$x_- = 1 - \frac{1}{\sqrt{2}}, \qquad x_+ = \frac{1}{3}\left(1 + \sqrt[3]{\frac{3}{4}\sqrt{78}+1} - \sqrt[3]{\frac{3}{4}\sqrt{78}-1}\right). \tag{24}$$

The roots of (24) are numerically equal to $x_- \cong 0.2929$ and $x_+ \cong 0.3966$. Solution $x_-$ corresponds to the values of the parameters

$$\beta_- = \frac{1}{2 - \sqrt{2}}, \qquad \lambda_- = 2 - \sqrt{2}, \tag{25}$$

i.e., they correspond to an internal mode with lowest frequency $\omega_-$, which turns into an instability mode at $\lambda_- \cong 0.5858$. For the localized mode with frequency $\omega_+$, numerical values of the parameters are, respectively, equal to $\beta_+ \cong 0.9180$ and $\lambda_+ \cong 1.089$.

Let us now carry out an analysis of the solutions of Eq. (21) in the limit of large $m$, i.e., for large-size clusters. It is obvious that, in the basic approximation, when the power term $x^m$ is negligibly small, $x = x_0 = 1/3$, and $\beta = \beta_0 = 4/3$. This case corresponds in fact to two remote Ising boundaries, the only localized mode of oscillation of which is the internal mode with squared frequency $\omega_0^2(\lambda) = 1 - \beta_0 \lambda$. It becomes an instability mode when $\lambda_0 = 1/\beta_0 = 3/4$ [3,10].

Consideration of the small values of $x^m$ leads to splitting of the limiting solution $x_0$ and the presence of two close roots in Eq. (21), which are given with good accuracy by the expression

$$x_{\pm} \approx \frac{1}{3} \pm 4 \left(\frac{1}{3}\right)^{m+2} . \qquad (26)$$

Accordingly, for the parameter $\beta$ and for the squared frequencies of internal modes, we obtain

$$\beta_{\pm} \approx \frac{4}{3} \mp 32 \left(\frac{1}{3}\right)^{m+2}, \qquad \omega_{\pm}^2 = 1 - \beta_{\pm} \lambda . \qquad (27)$$

Comparison of the numerical solution of Eq. (21) with formula (26) shows that, even for a cluster with $m = 3$, analytic expression is a very good approximation, which coincides with the roots of Eq. (21) with an accuracy of up to $0.5\%$. As a result, it turns out that the frequencies of internal modes and stability boundaries of clusters with $m \geq 3$ are determined explicitly by formulas (26) and (27), and, with increasing cluster size, its stability boundary with respect to the parameter $\lambda$ rapidly tends to the limiting value $\lambda_0 = 3/4$ according to the law

$$\lambda_{-} \approx \frac{3}{4} - 2 \left(\frac{1}{3}\right)^m . \qquad (28)$$

The resulting dependencies of the frequencies of the localized oscillations on the parameter $\lambda$ for nanoclusters with a number of spins from one to eight are shown in Fig. 1. It can be seen that the frequency dependence of $\omega_0(\lambda)$ for an isolated Ising boundary separates the frequency ranges of two modes with different symmetry. If we select the origin of coordinates so that it is located in the center of the cluster, then the frequency range below the curve will correspond to the even mode of oscillations $\omega_0(\lambda)$, and the odd mode is the range above this curve. The even mode describes antiphase oscillations of two domain boundaries, and the odd mode describes in-phase oscillations (Fig. 2). For considerably large distances between the boundaries of the oscillation mode, the difference and sum of the eigenfunctions of the spectral problem separated by the cluster size for an isolated Ising boundary correspond to each other. As the number of spins in the cluster decreases, i.e., when the Ising boundaries are approached, there is an increase in the distance between the frequency curves of these two modes from each other and, naturally, from the dividing limiting dependence for the isolated Ising boundary (Fig. 1). The fundamental difference between the modes is clearly manifested at the critical points $\lambda_{-}$ and $\lambda_{+}$, in which the modes soften and new static solutions appear in a bifurcational manner. The even mode added to the cluster (6) Fig. 1. with the distribution $(0, \pi, 0)$ leads to the appearance of a soliton-antisoliton configuration consisting of noncollinear boundaries with different signs, and the odd mode added to the cluster $(0, \pi, 2\pi)$ to soliton-soliton configuration forming a $360^0$ noncollinear domain wall.

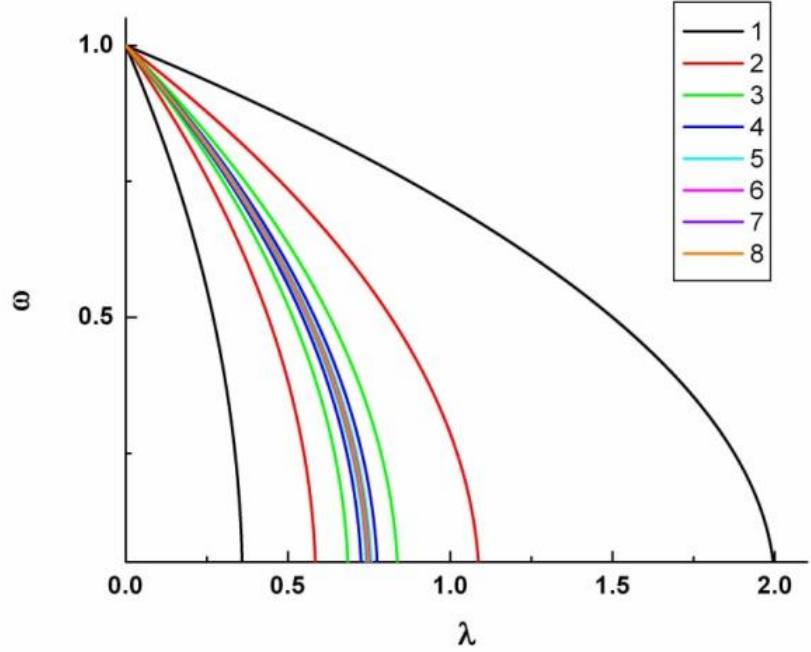

**Fig. 1.** *Frequencies of two localized oscillation modes of nano-clusters with number of spins from 1 to 8.*

The sequence of critical values $\lambda_-$ obtained above establishes the stability boundaries of clusters as their sizes increase. As can be seen in Fig. 1, for a ferromagnetic chain with given value of parameter $\lambda = \lambda_*$, nanoclusters with small number of spins for which $\lambda_- < \lambda_*$ will either be completely destroyed or transformed into dynamically unstable noncollinear soliton-antisoliton structures, while clusters with large number of particles for which $\lambda_- > \lambda_*$, remain practically unchanged. It turns out that the distribution function of localized oscillations with respect to frequencies has a condensation point at the oscillation frequency of the isolated Ising boundary $\omega_0(\lambda_*)$. Then, it should be expected that, in quasi-one-dimensional ferromagnets, whose exchange and anisotropy parameters are such that one- and two-spin clusters are unstable in them, the resonance peak of the MW field absorption will have only diffuse maximum at the frequency $\omega_0(\lambda_*)$. In the case of stable single- and double-spin nanoclusters, especially near the stability limit, local peaks, which are strongly spaced from the resonant maximum at frequency $\omega_0(\lambda_*)$ may appear in the absorption spectrum. Therefore, the detection of these peaks on the resonance curves does not only evidence the presence of spin nanoclusters, but also provides additional information on relationship between the exchange parameters and the anisotropy in quasi-one-dimensional ferromagnets.

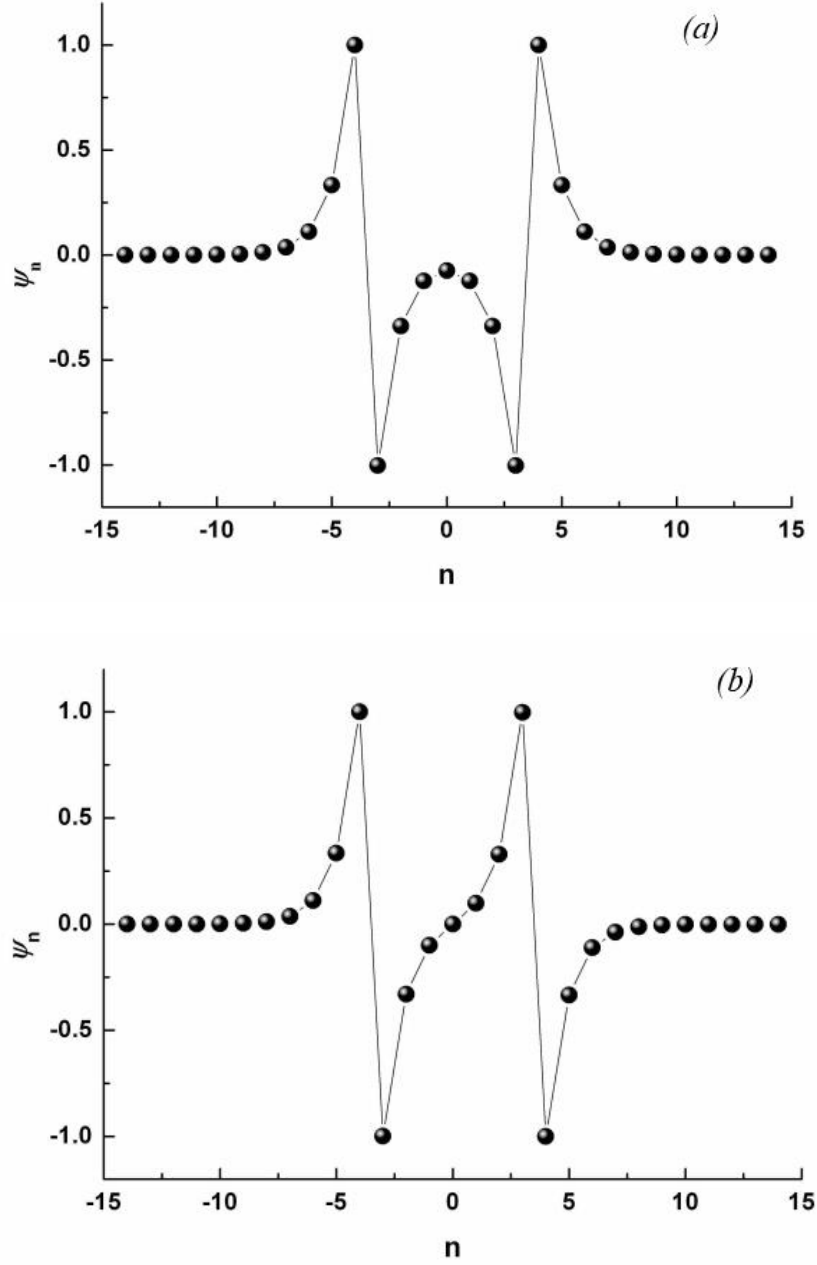

**Fig. 2.** *Localized oscillation modes of a cluster with 7 spins: (a) even oscillations mode corresponding to the minimum frequency; (b) odd oscillations mode.*

### 4. Scattering of spin waves in a ferromagnet with a spin cluster

The presence of stable nanoclusters in a ferromagnet has significant effect on the propagation of spin waves in it. To examine this, let us consider, within the framework of linearized equation (9), the following problem of scattering of continuous-spectrum waves on a spin cluster of arbitrary size. Assume that with $n = -\infty$ on a cluster with boundaries in nodes $n = j$ and $n = l$, a wave with single amplitude falls. Partially reflected, it passes through the cluster, and when $n \to \infty$, it has an amplitude whose squared modulus is equal to the spin-wave transmission coefficient

$$\psi_n = \exp(ikn) + A\exp(-ikn), \qquad n < j$$
$$\psi_n = B\exp(ikn) + C\exp(-ikn), \qquad j \le n \le l \qquad (29)$$
$$\psi_n = D\exp(ikn), \qquad n > l$$

Expressions (29) are exact solutions of Eq. (9) in the corresponding regions of the ferromagnetic chain, where quasi-wave vector $k$ is related to frequency $\omega$ according to the dispersion law (11). On the boundaries of the cluster and the external nodes adjacent to them, the equations for the amplitudes have the form

$$\psi_j - \psi_{j-2} - \mu\psi_{j-1} = 0,$$
$$\psi_{j-1} - \psi_{j+1} - \mu\psi_j = 0,$$
$$\psi_{l+1} - \psi_{l-1} - \mu\psi_l = 0, \qquad (30)$$
$$\psi_l - \psi_{l+2} - \mu\psi_{l+1} = 0.$$

Substituting solutions (29) into the system of Eq. (30), we obtain an inhomogeneous system of equations for determining the coefficients $A, B, C, D$. The solutions of this system of algebraic equations for the coefficients $A$ and $D$ are as follows:

$$A = e^{ik(l+j-2)} \frac{V^* e^{ikr} - V e^{-ikr}}{V^2 - e^{2ikr}}, \qquad (31)$$

$$D = \frac{e^{-2ik}}{\left(V^2 - e^{2ikr}\right)} \left(\frac{q - q^*}{|q|^2 - 1}\right)^2, \qquad (32)$$

In the formulas (31) and (32), the following notations are introduced:

$$V = \frac{q^2 - 1}{|q|^2 - 1}, \qquad q = \mu + e^{ik} = 2 - e^{-ik}, \qquad (33)$$

where $\mu = \mu(k)$ is given by the formula (11) and parameter $r$ is a difference of the numbers of boundary spins in the cluster $r \equiv l - j = m - 1$. It is obvious that moduli squares of the coefficients $A$ and $D$, i.e., reflection coefficient $R = |A|^2$ and transmission coefficient $T = |D|^2$, do not depend on the position of the cluster in the chain, but only on its size through the parameter $r$. It is not difficult to see that, as expected, the sum of these coefficients is equal to 1: $R + T = 1$. Noting also that $q - q^* = 2i\sin k$ and $|q|^2 - 1 = 4(1 - \cos k) = 2\mu$, the following representation is obtained for the reflection and transmission coefficients:

$$R == \frac{F^2(k)}{1 + F^2(k)}, \qquad T == \frac{1}{1 + F^2(k)}, \qquad (34)$$

$$F(k) = 8\,\mathrm{Im}\!\left(Ve^{-ikr}\right)\tan^2\frac{k}{2} = \cos^{-2}\frac{k}{2} \cdot \mathrm{Im}\!\left((q^2 - 1)e^{-ikr}\right). \qquad (35)$$

It follows from (34) and (35) that, when the function $F(k)$ becomes zero, the reflection coefficient $R = 0$, and transmission coefficient $T = 1$, that is, there is an effect analogous to the Ramsauer effect in quantum mechanics [15, 16]. In other words, there are such selected values of the quasi-wave vector $k$ for which the cluster is completely transparent for passage of spin waves. It is obvious that in the formula (35), the imaginary part of the complex expression $\text{Im}\left((q^2-1)e^{-ikr}\right)$ becomes zero when its argument $\chi = \pi p$, where $p$ is the integer. Noting that

$$q^2 - 1 = |q-1|\exp\left(i\frac{\pi - k}{2}\right) \cdot |q+1|\exp\left(i\arctan\left(\frac{\sin k}{3 - \cos k}\right)\right), \qquad (36)$$

we obtain equation for finding values of the quasi-wave vector for which $T = 1$,

$$k\left(m - \frac{1}{2}\right) - \arctan\left(\frac{\sin k}{3 - \cos k}\right) = \pi\left(p - \frac{1}{2}\right), \qquad p = 1,..,m-1. \qquad (37)$$

It is easy to see graphically that the number of roots of this equation inside the interval $[0, \pi]$ is equal to $m-1$, because $p = m$ corresponds to the interval margin $k = k_B = \pi$. For large values of $m$ formula (37) yields an explicit analytic expression for unknown roots

$$k = k_p + \frac{1}{m-1}\arctan\left(\frac{\sin k_p}{3 - \cos k_p}\right), \qquad k_p = \frac{\pi}{m - \frac{1}{2}}\left(p - \frac{1}{2}\right). \qquad (38)$$

It turns out to be a very good approximation not only for large clusters, but even for a cluster with two spins, where, as comparison with the numerical result shows, accuracy for the smallest roots is of the order of a percent; for the rest of the roots, it is of the order of tenths and hundredths of a percent. Note that in the case of a single-spin cluster, the transparency coefficient has only one maximum $T(0) = 1$, and it corresponds to $F(0) = 0$, which is ensured not by the zero value of the argument $\chi$, but by the zero value of the module $|q-1|$ at this point in expression (36). Figure 3 shows the dependence of the transmission coefficient on the quasi-wave vector for nanoclusters of 5 and 12 particles. It may be seen that, in the long-wavelength limit, spin waves pass almost unobstructed through nanoclusters, while short-wave oscillations almost completely reflect from clusters long before boundary of the Brillouin zone. In the intermediate region, maxima of total transparency are observed, whose number, together with the long-wave maximum, is equal to number of spins in the cluster. In this case, the resonance peaks become narrower as they appear with increasing cluster size. Obviously, when passing spin wave packets, clusters behave like frequency filters. It should be expected that, in a chain with distributed clusters, short-wave excitations will be localized between clusters, and only long-wave oscillations will pass through the entire system. However, if an isolated spin cluster is excited or created by a local field in a quasi-one-dimensional ferromagnet or metamaterial, then its size, i.e., the number of spins that form it, may be determined according to the results of spectral analysis of the wave transmitted through it.

To verify the obtained theoretical results, numerical simulation of the dynamics of the ferromagnetic chain containing clusters with two, three, and four spins was carried out within the framework of the nonlinear Takeno-Homma equation (3). The results are shown in Fig. 4. The value of parameter $\lambda$ was selected to be equal to $\lambda_* = 0.7$, at which the two-spin cluster is obviously unstable, the cluster of four spins is still stable, and the cluster of three spins has already lost stability (see Fig. 1). In the initial distribution of nanoclusters with configuration $(0, \pi, 0)$, the two-spin cluster was located at the origin of the coordinates between the three-spin and four-spin clusters, which were strictly in an equilibrium state, while the two-spin cluster had an additive to $\pi$ equal to $10^{-4}$, for both amplitudes. Therefore, as follows from the analytical results obtained, the two-spin cluster immediately collapsed completely. At the nonlinear stage of evolution, a strongly localized discrete breather was formed from it, and packets of spin waves began to propagate from it in both directions. As can be seen in Fig. 4, long-wavelength components of the packets passed through a three-spin and four-spin cluster; however, the short-wave part was trapped between them. As a result of perturbations, the three-spin cluster also lost stability, turning into a soliton-anti-soliton discrete configuration, which turned out to be dynamically unstable, leading to essentially nonlinear oscillations of its amplitudes. Further, Fig. 4 shows that the four-spin cluster remains stable, experiencing only small fluctuations in amplitude despite collisions with spin waves. Subsequently, the discrete breather starts to move toward the four-spin cluster and is reflected from it. At a late stage of evolution, it passes a three-spin dynamic complex, finally destroying it; this results in the formation of another discrete breather.

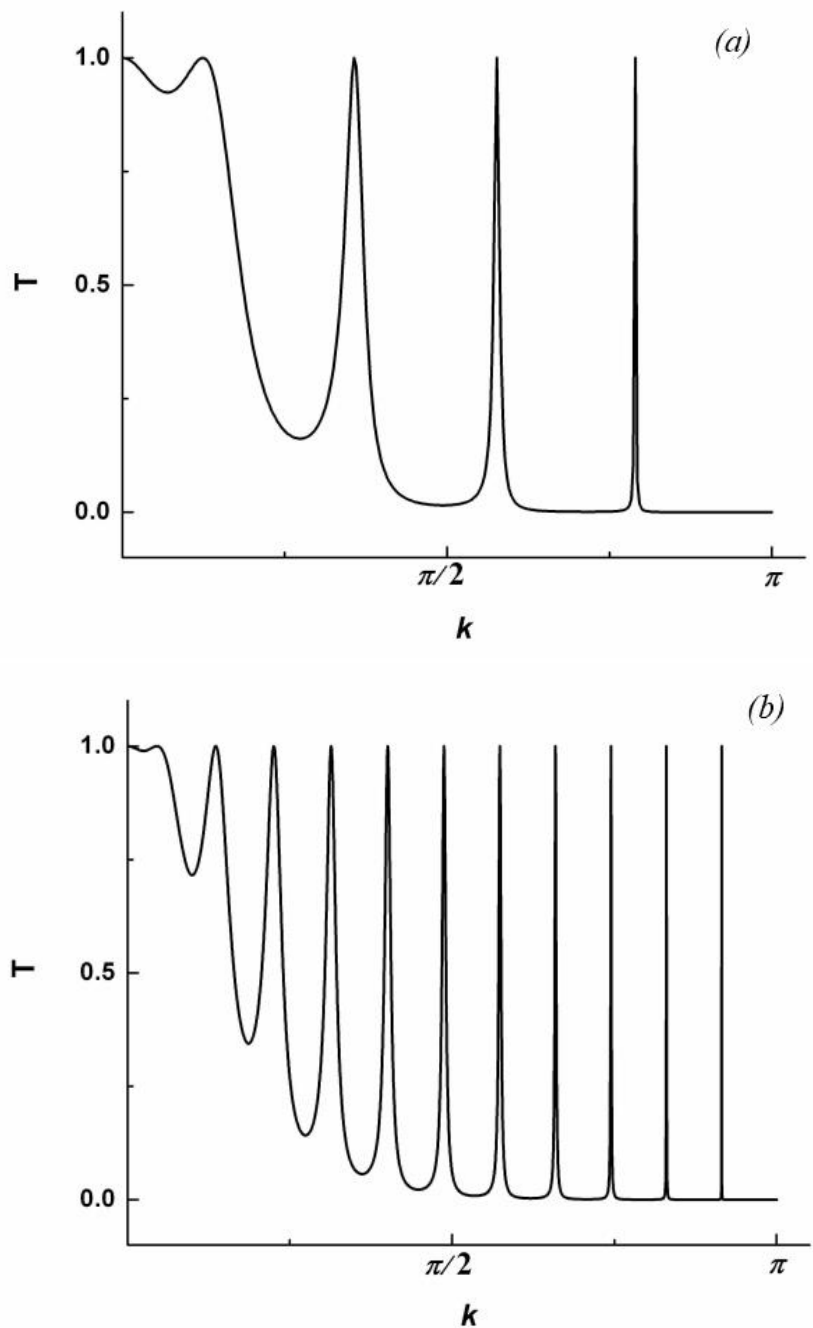

**Fig. 3.** *Transmission coefficient $T$ of spin waves passed through nanoclusters as a function of quasi-wave vector $k$ : (a) for a cluster of 5 spins; (b) for a cluster of 12 spins.*

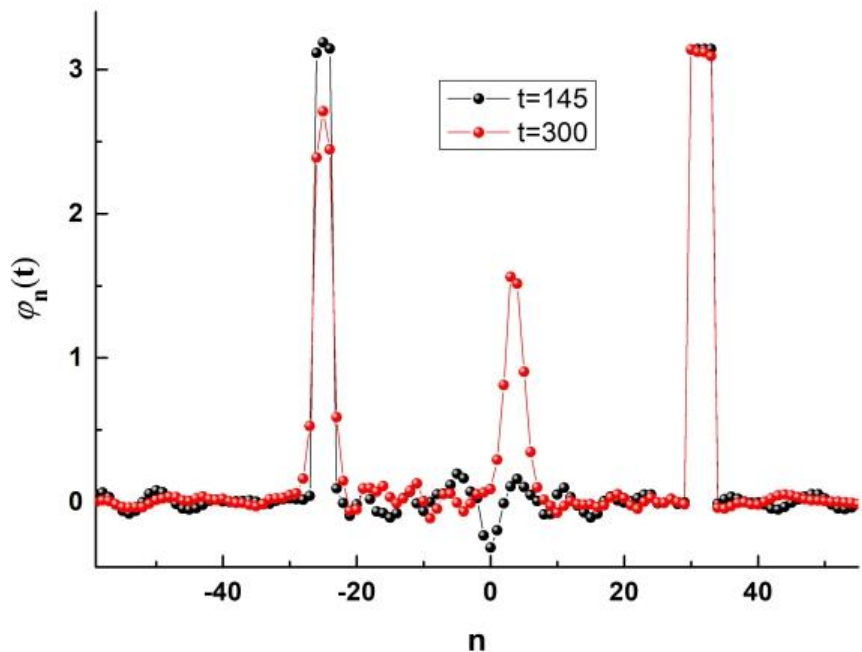

**Fig. 4.** *Results of modeling of the dynamics of a ferromagnetic chain containing clusters with 2, 3, and 4 spins within the framework of a nonlinear THE.*

### 5. Conclusion

In this paper, the spectral problem for localized excitations and the problem of the scattering of spin waves in an anisotropic ferromagnetic chain containing spin clusters are solved exactly within the framework of the discrete Takeno-Homma model. Using the method of local perturbations, frequencies of localized oscillations of a spin cluster of arbitrary size are found which describe the antiphase and in-phase oscillations of its discrete domain walls. Critical values of the exchange-to-anisotropy ratio are found under which the lowest frequency mode is "softened" and the instability of clusters results. The problem of spin waves scattering on arbitrary size clusters is solved and it is shown that there is a dimensional effect of resonant full wave propagation with discrete set of values of the quasi-wave vector, similar to the Ramsauer effect in quantum mechanics. The observed features of the dynamics of the magnetic excitations may be observed not only in ferromagnets and antiferromagnets [10], but also in other systems of interacting planar rotators [9, 17]. The creation of metamaterial consisting of interacting magnetic molecules chains and allowing local excitation of magnetic nanoclusters is particularly attractive. Such metamaterials that possess the abovementioned properties may be used as frequency filters, with application as an elementary base of computer memory in modern nanotechnologies.

The work was supported by the FFI of the NAS of Ukraine (Grant No. 1.4.10.26.4/F26-4) and NAS of Ukraine (Grant No. 4/17-H) and implemented using the Grid Cluster computing resources of the B.I. Verkin ILTPT of the NAS of Ukraine.